# A Growth Model for DNA Evolution

M. de Sousa Vieira[*][†][&] and H. J. Herrmann[*][%]

[*] *Hochstleistungsrechenzentrum Supercomputing Center, Kernforschungsanlage, D-52425 Jülich, Germany.*

[†] *Departamento de Física, Universidade Federal do Rio Grande do Norte, Campus Universitário, Natal, RN 59072-970, Brazil.*

[%] *P. M. M. H. (U. R. A. CNRS 857) E. S. P. C. I., 10 rue Vauquelin, 75231 Paris Cedex 05, France.*

## Abstract

A simple growth model for DNA evolution is introduced which is analytically solvable and reproduces the observed statistical behavior of real sequences.

Typeset using REVTEX

The study of fractal structures in DNA chains is a subject of great interest nowadays. Two exciting reports of the existence of long-range correlations, which implies fractal structure, in DNA sequences have been made recently by Li [1] and Peng et al. [2,3].

DNA sequences are formed of four different nucleotides: Adenine, Guanine, Thymine and Cytosine, which are represented by the letters A, G, T, and C, respectively. The first two are called purines and last ones pyrimidines. Roughly, half of the nucleotides in a chromosome are pyrimidines and half are purines.

It is observed that the length of the DNA chain of evolved species, such as mammals, is much longer than the DNA of simple organisms, like bacteria [4]. Thus, DNA sequences are in a dynamical state, with its length increasing during evolution. Often the chain increases by gene duplication. Extra copies of the gene are stored in the same sequence as the original copy. Consequently, DNA sequences of evolved species usually have repetitive patterns [4]. However, the copy is not always identical to the original gene. During the process of duplication a mutation can occur, and more complex functions can appear in the evolving species. In our view, the reason for this is that it causes less disruption to normal functioning to keep a copy of the unmutated gene, because if the mutated gene does not work perfectly, the first copy can continue to function. Sometimes, the repeated segments are next to each other, sometimes they are separated. The length of these repeated segments varies greatly; some of them are short, others are as long as the entire gene. A set of genes descended by duplication and variation of some ancestral gene is called a *gene family* [4]. Copy numbers in a given family can vary from two to millions.

It has been found that the content of repetitive DNA tends to increase with increasing chain length. Another important observation is that only a small fraction of the DNA chain is active, namely, codes for protein. The rest of it apparently has no function [4]. In humans the percentage of noncoding DNA can be larger than 90%.

Peng et al. [2] analyzed the statistical properties of real DNA chains by constructing a graphical representation of the sequence. Starting from the origin, a walker moves either "up" [$u(i) = +1$] if a pyrimidine (C or T) occurs at a position $i$ or the walker steps "down" $[u(i) = -1]$ if a purine (A or G) occurs at position $i$. Next, they calculate the cumulative displacement $y(l)$ of the walker after $l$ steps, which is the sum of the unit steps $u(i)$ at each position $i$. Thus, $y(l) = \sum_{i=1,l} u(i)$. A fractal landscape is constructed by plotting $y(l)$ versus $l$.

The walk is characterized by the root mean square fluctuation $F(l)$ about the average displacement; $F(l)$ is defined in terms of the difference between the average of the square and the square of the average $F^2(l) \equiv \overline{[\Delta y(l) - \overline{\Delta y(l)}]^2} = \overline{[\Delta y(l)]^2} - \overline{\Delta y(l)}^2$, of a quantity $\Delta y(l)$ defined by $\Delta y(l) = y(l_0+l) - y(l_0)$. Here the bars indicate an average over all positions $l_0$ in the sequence.

Three possibilities exist for $F(l)$: (i) if the walk is random, then $F(l) \sim l^{1/2}$. (ii) If there are local correlations extending up to a characteristic range, then asymptotically $F(l) \sim l^{1/2}$ would be unchanged from the purely random. (iii) If there is no characteristic length (i.e., if the correlation were "infinite-range"), then $F(l) \sim l^\alpha$ with $\alpha \neq 1/2$.

In Refs. [2,5,6] was found that $\alpha > 0.5$ for noncoding segments and $\alpha = 0.5$ for coding segments. On the other hand, in Ref. [7] it was reported that in most cases $\log F(l)$ against $\log l$ is not a straight line. It was found nonlinear curves both for coding segments and those containing noncoding regions with a local slope larger than 0.5. According to these results, a well-defined fractal power exponent $\alpha$ does not seem to exist for DNA sequences. The behavior found has been that, for small $l$, $\alpha$ is approximately 0.5 then increases monotonically to a maximum value. Beyond this, the statistics have not been good enough to allow a conclusion about the asymptotic value of $\alpha$ in the limit of very large $l$. However, there are clear indications in [3] that $\alpha$ decreases after attaining a maximum, although this behavior was not investigated in detail for real sequences.

We introduce here a simple iterative model for gene evolution, which is analytically solvable and reproduces the behavior of the exponent $\alpha$ found in [3,7] for real DNA sequences of intermediate size. The model incorporates basic features of DNA evolution, that is, sequence elongation due to gene duplication and mutations. When very large chains are analyzed in our model, the exponent $\alpha$ is back to the trivial value 0.5. Therefore, our model suggests that the nontrivial value found for $\alpha$ in [2] is probably due to finite size effects. We are aware of two other models [1,8] for DNA evolution, but in both of them the critical exponents do not go back to the trivial values of a random sequence as the chain gets very large.

Our dynamical model for DNA evolution is the following: we start with a coding sequence which consists of $g$ genes of equal length. Each gene has $n$ nucleotides, which consists of pyrimidines ($u = +1$) and purines ($u = -1$) randomly distributed in a proportion of 50% each. In the simulated evolutionary process we choose at random a gene of our original sequence and in that gene we choose at random one of its nucleotides. Then, we change (mutate) this nucleotide from a pyrimidine to a purine or vice-versa. In real DNA a mutation in a single nucleotide is called a *point mutation* [4]. A copy of the chosen gene before the mutation is added at the end of the chain. This old gene becomes a noncoding segment, and it is not further modified. The genes that can mutate are always the first $g$ genes, which are the coding DNA in our model. We iterate this process $N_i$ times ($N_i \gg 1$). At the end, our chain will consist of a "head" of $g$ coding genes and a big "tail" of $N_i$ noncoding segments. The final length of the chain is $L = n(g + N_i)$.

It is clear that our model is a simplified version of what is found in real sequences. Some of the features that our model misses are the following: In real DNA, (a) the genes do not have the same size; (b) not all sites of a gene have identical probability of mutation; (c) the distribution of purines and pyrimidines does not have a random uniform distribution; (d) a point mutation can be from a purine (pyrimidine) to the other kind of purine (pyrimidine); (e) noncoding segments are interdispersed with coding ones, and (f) mutations can also occur via deletions and insertions of genetic material. Our preliminary results show that if we introduce in the model the features (a) to (e) the results shown here are changed only quantitatively (this will be reported elsewhere). We have not studied yet feature (f).

In Fig. 1 we show a fractal landscape of our growth model, given by $y(l)$ versus $l$, for the parameter values $g = 3$, $n = 10$ and $N_i = 10^4$. The landscape presents a jagged contour, as found in real DNA sequences [2], indicating the possible existence of long-range correlations.

The mean square fluctuation $F(l)$ is related to the autocorrelation function $C(l) = \overline{u(l_0)u(l_0+l)} - \overline{u(l_0)}^2$ through the relation $F^2(l) = \sum_{i=1,l}\sum_{j=1,l} C(j-i)$. In our model we can calculate analytically $C(l)$ and consequently $F(l)$ and $\alpha(l)$. Let us first analyze the term $\overline{u(l_0)u(l_0+l)}$ that appears in $C(l)$. Since in our original random sequence of coding segments the nucleotides are not correlated to each other, this implies that correlations exist only for sites which are at a distance that is a multiple of $n$. When this is not the case, $\overline{u(l_0)u(l_0+l)}$ will be an average of $L-l$ random terms with values $\pm 1$. Because of the central limit theorem, this term will be a gaussian random variable with zero mean and variance equal to $1/\sqrt{L-l}$. If we consider many realizations of the growth process, the ensemble average of this random variable will be zero. Consequently, in this limit $\overline{u(l_0)u(l_0+l)} = 0$ for $l \neq mn$ ($m = 1, 2, 3, ...$) and $l \neq 0$. For $l = 0$ one obviously has $\overline{u(l_0)u(l_0)} = 1$.

To find $\overline{u(l_0)u(l_0+l)}$ for the case in which $l = mn$ will consider first the case of a single gene, $g = 1$, with $n > 1$. When the gene is duplicated a mutation occurs and we have with probability $p_1 = 1/n$ that $u(l_0+n)$ has a different sign as the one of $u(l_0)$, and with probability $1 - p_1$ that their signs are the same. This is valid for any $l_0$. Then $A \equiv \overline{u(l_o)u(l_o+n)} = 1 - 2p_1 = (n-2)/n$, since $u(l_o) = \pm 1$. We have the probability $p_2 = 2(n-1)/n^2$ for $u(l_o+2n)$ being different from $u(l_0)$ and $1-p_2$ for them being identical. Consequently, $\overline{u(l_o)u(l_o+2n)} = (n-2)^2/n^2 = A^2$. We can continue the process and easily find that for the case of a single gene $\overline{u(l_o)u(l_o+mn)} = A^m$.

For $g > 1$ we have to consider that $u(l_0+n)$ has the probability $1/g$ of being originated from the same gene as $u(l_0)$. Consequently, it has probability $p_1 = 1/(ng)$ of being different from $u(l_0)$, and $(n-1)/(ng)$ of being identical. Note that here the sum of these two probabilities is not one, since there exists also the probability $(g-1)/g$ that $u(l_0+n)$ is originated from a different gene, and this gives on average a vanishing contribution for the term $\overline{u(l_o)u(l_o+n)}$. In this way, $\overline{u(l_o)u(l_o+n)} = A/g$. Between $u(l_0+n)$ and $u(l_0+2n)$ we see the following differences: there are $g^2$ possible different combinations for the gene that contains $u(l_0+2n)$. One of these combinations corresponds to a gene that is mutated twice and $(1-g)$ possibilities that a gene is mutated once. This will result in $\overline{u(l_o)u(l_o+2n)} =$ $(A^2 + (g-1)A)/g$. It is not difficult to find that there are $g^3$ possible different combinations for the gene that contains $u(l_0 + 3n)$. One of these possibilities corresponds to a gene that is mutated three times, $2(g-1)$ that a gene is mutated twice and $(g-1)^2$ the gene is mutated once. Consequently, $\overline{u(l_0)u(l_0 + 3n)} = A[A + g - 1]^2/g^3$. In this way, we can continue the process and find that $\overline{u(l_0)u(l_0 + mn)} = A[A + g - 1]^{m-1}/g^m$. These results are valid for a large number of realizations of the growth process. However, note that the ensemble average is only important if the chain is not very large, because otherwise the spatial average will already give good statistical results.

To determine $\overline{u(l_0)}^2$ in the expression for $C(l)$, we note that even if the number of pyrimidines is different from the number of purines in the initial chain of coding segments, after few iterations the probability of having $u = -1$ will be identical to the probability of having $u = +1$. The easiest way to show this is for the case of a single gene. Consider a nucleotide with $u(l_0) = +1$. The probability $P_+^m$ of having $u(l_0 + mn) = u(l_0 + (m-1)n)$ is $P_+^m = \frac{n-1}{n} P_+^{m-1} + \frac{1}{n}(1 - P_+^{m-1})$, with $P_+^0 = 1$. From this recursive relation we find $P_+^m = (1 + A + A^2 + ... + A^m)/n$. The sum of this geometric series converges to $1/2$. Therefore, for large $m$, the probability of finding a nucleotide with $u(l_0 + mn) = +1$, as it was in the beginning of the iteration process, is identical to the probability of finding $u(l_0 + mn) = -1$. Since by definition $\overline{u(l)} = \frac{1}{L}\sum_{l_0=1,L} u(l_0)$ and $u(l_0)$ has probability $1/2$ of being $+1$ and $1/2$ of being $-1$, it turns out that $\sum_{l_0=1,L} u(l_0)$ evolves as a symmetric random walker in one-dimension. It is well known that such a walker diffuses away from the origin with average distance of $\sqrt{L}$. Consequently, $\overline{u(l)} \approx 1/\sqrt{L}$.

Combining the results found above, we have the following expression for the autocorrelation function

$$C(l, L) = \begin{cases} 1 - 1/L, & \text{if } l = 0, \\ A[A + g - 1]^{m-1}/g^m - 1/L, & \text{if } l = mn, \\ -1/L, & \text{otherwise.} \end{cases} \quad (1)$$

This shows that the autocorrelation function is a function of $l$ and of the length $L$ of the chain. The above equation does not apply when $n = 1$ and $g = 1$, where $C(l, L)$ is a period-two function. Also, for the case $n = 1$ and $g > 1$ all sites are uncorrelated to each other, which results in $\overline{u(l_o)u(l_o + l)} = 0$ when $l \neq 0$. Finally, the case $n = 2$ is not interesting since it results in $A = 0$. Therefore, we will concentrate our attention to cases in which $n > 2$.

From Eq. (1) we can obtain the correlation length $\tilde{l}$, which is determined by the smallest value of $l$, with $l = mn$, satisfying $C(l, L) = 0$. Namely, $\tilde{l} = \tilde{m}n$, where $\tilde{m}$ is given by

$$\tilde{m} = \log\left[\frac{A + g - 1}{LA}\right] \bigg/ \log\left[\frac{A + g - 1}{g}\right]. \tag{2}$$

Since $F(l, L)$ can be obtained from $C(l, L)$, and if we take into consideration that $C(-l, L) = C(l, L)$, we find

$$F(l, L) = \sqrt{l[1 + 2\sum_{j=1,l} C(j, L)] - 2\sum_{j=1,l} jC(j, L)}. \tag{3}$$

As in [3], we calculate the local slope $\alpha(l_i, L)$ of $\log_{10} F(l_i, L)$ versus $\log_{10} l_i$, given by

$$\alpha(l_i, L) = \frac{\log_{10} F(l_{i+1}, L) - \log_{10} F(l_i, L)}{\log_{10} l_{i+1} - \log_{10} l_i}. \tag{4}$$

If we take $l_{i+1} = l_i + k$, then

$$\alpha(l_i, L) = \log_{10}\left\{1 + \frac{k\left[1 + 2\sum_{j=1}^{l_i+k} C(j, L)\right] + 2\sum_{j=l_i+1}^{l_i+k}(l_i - j)C(j, L)}{l_i + 2\sum_{j=1}^{l_i}(l_i - j)C(j, L)}\right\} \bigg/ 2\log_{10}\left\{1 + \frac{k}{l_i}\right\}. \tag{5}$$

For $l < ng$ or large $l$, we have $C(l, L)$ of the order of $-1/L$, this implies that for very large chains Eq. (5) gives $\alpha \approx 0.5$. For intermediate values of $l$ we have that $\alpha$ presents a spiked behavior at $l = mn$, as a consequence of the spikes present in the correlation function (see Eq. 1). We observe that if we average $\alpha(l, L)$ for $n$ consecutive values of $l$, with $l \geq ng$, the curve $\alpha$ versus $\log l$ becomes smooth. Another equivalent way to smooth $\alpha(l, L)$ is to take $k = n$ and $l_i$ a multiple of $n/2$. This is the method we use here.

We show in Fig. 2(a) $\log_{10} F(l, L)$ versus $\log_{10} l$ for $n = 10$, $g = 3$ (solid), $n = 10$, $g = 10$ (dashed) and $n = 6$, $g = 3$ (dotted). In all the three cases, $N_i = 10^4$. As in real DNA sequences [3,7] we do not find a straight line. Therefore, also in our growth model a well-defined fractal power exponent $\alpha$ does not exist. In Fig. 2(b) we show $\alpha$ versus $\log l$ associated with the curves shown in Fig. 2(a). We see that the local slope $\alpha$ starts from 1/2, then increases to a maximum value, and for large $l$ decreases again to 1/2. For these parameters values we find from Eq. (2) $\log_{10} \tilde{l} \approx 3.17$, 3.64, 2.69, respectively, which correctly mark the region in which $\alpha(l, L)$ returns to the trivial value 1/2. We see that, for small and large $l$ the exponent $\alpha$ is the one of a random sequence. Our results for $n = 10$, $g = 10$ and $n = 10$, $g = 3$ are in good agreement with Figs. 4 and 8 of [3]. However, in those figures the statistics for large $l$ are not as good as in our case, and $\alpha$ presents considerable fluctuations. The results obtained from our analytical expressions were compared with the ones obtained by direct simulation of the growth process, and they were in excellent agreement with each other. Our picture indicates that for large $l$, $\alpha(l, L) = 1/2$ in real DNA sequences if chains sufficiently big were analyzed in [3].

We also analyzed the Fourier spectrum of the numerical sequence obtained through the random walk method described above. We see at large frequencies a white noise component (with peaks located at $f = 1/n$ and its respective harmonics, reflecting the presence of the spikes at $l = mn$ in the autocorrelation function). For intermediate frequencies we find $1/f$ noise and at small frequencies the power spectrum again flattens off. This can be observed only if the chain is very large. If this is not the case, only the $1/f$ and the white noise regions (at large frequencies) are observed (more details will be published elsewhere). This is exactly what has been seen in real DNA sequences [1,9]. By studying very large sequences, we are able to show that the $1/f$ noise in real DNA chains is apparently only a transient behavior observed in chains that are not sufficiently large.

In conclusion, we have introduced a growth model for DNA evolution which incorporates basic features of a DNA growth process, that is, sequence elongation due to gene duplication and mutations. Our model gives the same statistical results of real sequences investigated in the literature, that is, the apparent existence of long-range correlations. However, when larger chains are studied, we see that this is only a transient behavior valid for chains not sufficiently big. From our studies we suggest that a possible way to find the asymptotic value of $\alpha$ in real sequences, eliminating the statistical fluctuations, is to divide the chain in big segments and study the ensemble average of this exponent.

## ACKNOWLEDGMENTS

MSV thanks the Alexander von Humboldt Fundation and the Brazilian Agency CNPq for financial support. We thank W. Li, C.-K. Peng and H. E. Stanley for enlightening correspondence.

FIGURES

FIG. 1. Landscape (walk) representation of our model, with $n = 10$, $g = 3$ and $N_i = 10^4$. As in [7], for a more convenient representation, we removed the trend of the landscape in such a way that the end point has the same vertical displacement as the starting point.

FIG. 2. (a) $\log_{10} F(l, L)$ versus $\log_{10} l$. (b) $\alpha(l, L)$ versus $\log_{10} l$, for $n = 10$, $g = 3$ (solid), $n = 10$, $g = 10$ (dashed) and $n = 6$, $g = 3$ (dotted) with $N_i = 10^4$.

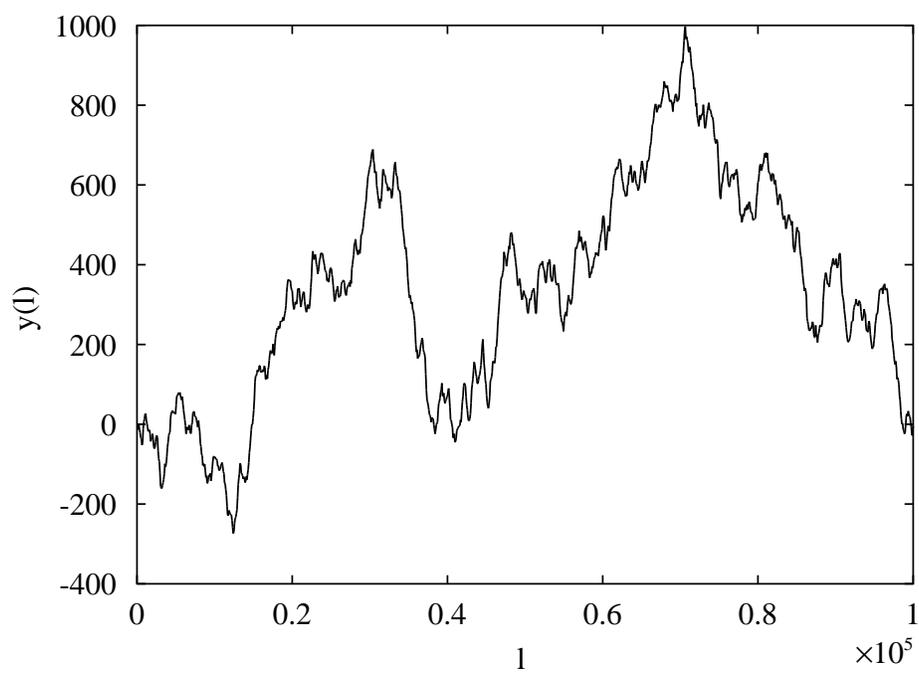

Fig. 1

Fig. 2(a)

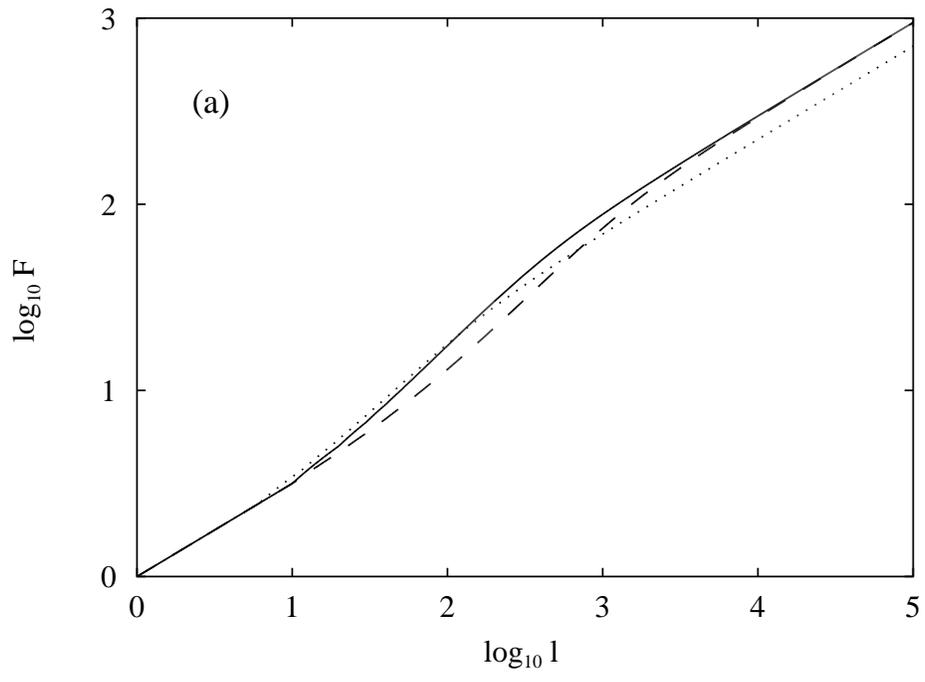

Fig. 2(b)

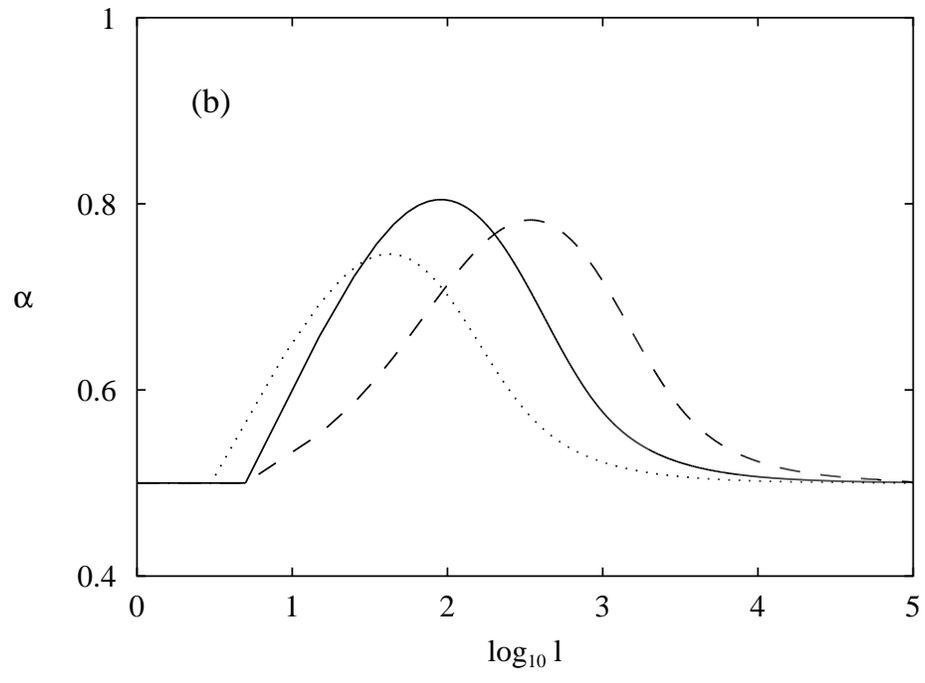